# Visualization of Defect Electronic States in Layered Semiconductor CrSBr

*Jonathan Brunette[1], Joost Aretz[2], Kiyoung Jo[3], Aljoscha Soll[4], Zdeněk Sofer[4], Malte Rösner[2,5], Nathan Guisinger[6], Adina Luican-Mayer[1,3,7*]*

[1]Department of Physics, University of Ottawa, Ottawa, K1N 6N5, Canada

[2]Institute for Molecules and Materials, Radboud University, Nijmegen, 6525 AJ, The Netherlands

[3]Material Science Division, Argonne National Laboratory, 60439, Lemont, Illinois, USA

[4]Department of Inorganic Chemistry, University of Chemistry and Technology Prague, Technicka 5, 166 28 Prague 6, Czech Republic

[5]Faculty of Physics, Bielefeld University, 33615. Bielefeld, Germany

[6]Center for Nanoscale Materials, Argonne National Laboratory, 60439, Lemont, Illinois, USA

[7]University of Illinois, 60607, Chicago, Illinois, USA

Chromium sulfur bromide (CrSBr) is a layered magnetic semiconducting material combining a rich magnetic phase diagram with axis-dependent electronic and optical properties. While defects in CrSBr have been shown to affect magnetic order and excitonic responses, their microscopic nature, atomic structure, and electronic properties are not yet fully understood. In this work, we use scanning tunneling microscopy/spectroscopy (STM/STS) to explore the structure and electronic signatures of two prominent defects in bulk CrSBr. Their structure reflects the symmetries of the underlying lattice, with electronic features near the valence band edge. By comparing experimental data with *ab initio* simulated STM images, we infer that a common defect corresponds to a b-axis-aligned double sulfur vacancy, in line with findings from a recent growth analysis study. This result advances our understanding of the role of intrinsic defects in shaping the electronic structure of CrSBr.

Defects play an important role in determining physical, chemical, and optical properties of materials and are inherently present in all real systems. For example, vacancies, anti-sites, charged impurities and grain boundaries can serve as scattering centers which influence carrier mobility and can introduce states within the band gap of semiconducting materials.[1] These effects can be particularly pronounced in two-dimensional (2D) systems, where the reduced dimensionality and high surface-to-volume ratios potentially enhance the density of defects. In semiconducting transition metal dichalcogenides, defect-induced electronic states affect the exciton binding energies, effectively acting as trapping and localization centers, governing exciton recombination and diffusion. This can lead to quenched and defect-bound photoluminescence and altered exciton lifetimes.[2–4] As a result, defect engineering is a pathway to tailor material functionalities.[1,5–8]

A recent addition to the family of layered semiconductors is CrSBr, an air-stable magnetic semiconductor with intrinsic long-range magnetism, quasi one-dimensional behavior, and intriguing magneto-optical properties, enabling new ways to couple electronic and spin degrees of freedom.[9–15] Its anisotropy leads to pronounced polarization-dependent optical transitions sensitive to layer thickness and strain.[16] In addition, since CrSBr features type-A antiferromagnetic ordering below its Néel temperature, strong coupling between the electronic and magnetic degrees of freedom enables magnetic field-dependent optical responses, where excitonic features persist down to the few layers limit.[9,10,17–21] In such a material, defects can change local charge density, disrupt magnetic order, or pin spins, directly modifying electrical conductivity, magnetism, and their coupling. Both theory and experimental works found that defects such as vacancies, anti-sites and interstitial atoms in CrSBr can contribute to n-type conductivity.[22] Vibrational spectroscopies revealed that the increased concentration of defects in CrSBr can lead to magnetic phases more robust to temperature.[23]

The defects in CrSBr have been explored using electron microscopy, where HAADF-STEM of bilayer CrSBr revealed surface Br vacancies to be the predominant defect type compared with S and Cr vacancies, in agreement with calculations of their respective formation and binding energies.[22–24] Theoretical models proposed that the presence of Cr interstitial atoms can lower the formation energy of Cr vacancies. Additionally, the presence of Br, S and Cr vacancies can further reduce the formation energy of adjacent defects, giving rise to defect complexes that can span over multiple unit cells.[24] A recent chemical vapor transport growth study, supported by thermodynamic and first-principles modelling, furthermore showed that certain defect concentrations can be significantly reduced by sulfur- and bromine-rich growth conditions, pointing towards prominent sulfur vacancies.[25] Despite substantial theoretical efforts to predict the energetic and electronic properties of these defects, direct experimental characterization of their electronic structure remains limited. Here, we employ STM to directly resolve the electronic structure of individual defects in CrSBr with atomic-scale resolution.

In Figure 1a, we show the rectangular crystal structure of CrSBr, with Br atoms located at the surface. When viewed in plane along the a-axis (Fig 1a, top), the Br atoms are bonded to two separate Cr atoms and aligned with the S atom on the c axis (Fig 1a, bottom).

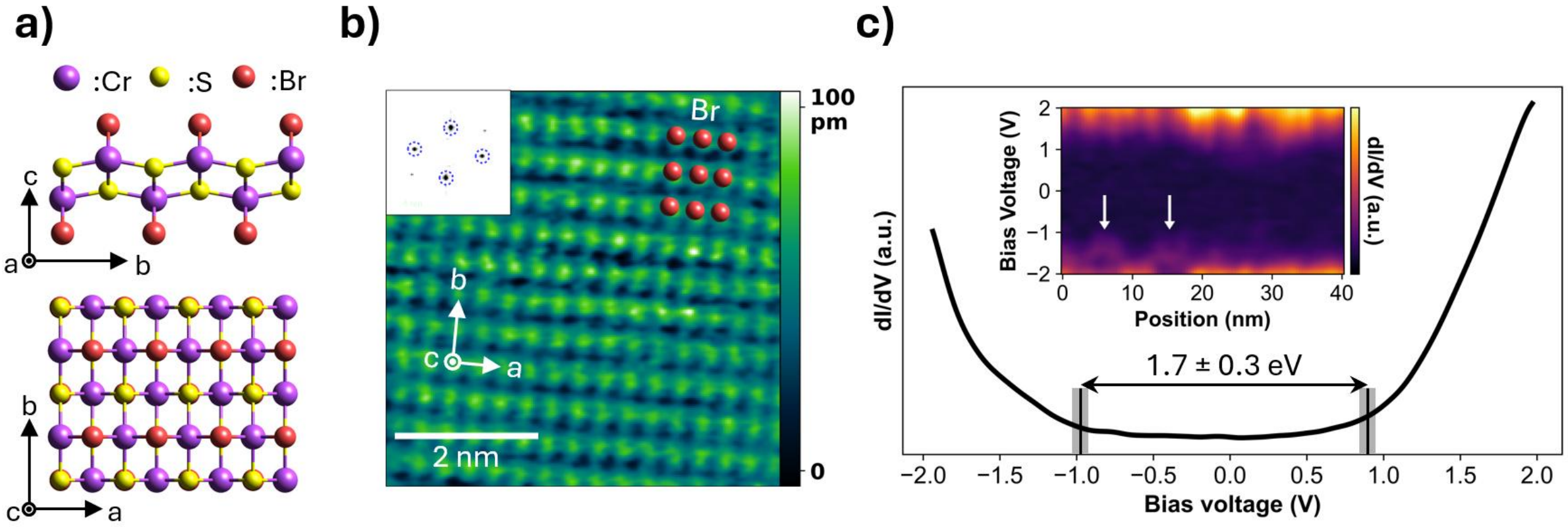

**Figure 1.** a) CrSBr crystal structure. b) STM topographic image of bulk CrSBr showing atomic sites of Br (VB = -1 V, IS = -50 pA). Inset: 2D FFT of the topographic image. c) STS of bulk CrSBr averaged over 40 nm (inset) (VB = -2 V, IS = -50 pA). Inset: Intensity map of all spectra corresponding to the average. All measurements were taken at 211K.

A scanning tunneling micrograph resolving individual atoms of bulk CrSBr is presented in Figure 1b, where the rectangular symmetry is apparent along the a and b directions. The symmetry is further observed in the 2D FFT of the topographic map, shown in the inset. From the atomically resolved topographs, we obtain lattice parameters $|\hat{a}|= 3.6 \pm 0.2$ Å, $|\hat{b}|= 4.6 \pm 0.2$ Å, and $\theta=89.7 \pm 0.5^{o}$, in agreement with existing literature.[26]

We performed scanning tunneling spectroscopy along a 40 nm line. The average spectrum in Figure 1c shows a band gap of $1.7 \pm 0.3$ eV, in agreement with existing measurements and theoretical predictions for bulk CrSBr.[27,28] In the inset of Fig.1c, we plot an intensity color map of the dI/dV spectra along this 40nm line, where we observe the presence of shoulder-like features near the edge of the valence band (marked with arrows), which we associate to the presence of defects.

When we zoom out across the surface, as shown in the topographic map of Figure 2a, we find signatures of nanoscale defects. From their qualitative features, we observe two types of defects, labeled as type 1 and type 2. From this map, we can extract defect densities of ~$2.4*10^{11}$ and ~$2.3*10^{11}$ $cm^{-2}$ for the respective defect types, in agreement with previous studies on CrSBr.[23,25] When compared to other semiconducting 2D materials such as transition metal dichalcogenides, this falls within their typical reported defect densities of $10^{10}$-$10^{14}$ $cm^{-2}$.[29]

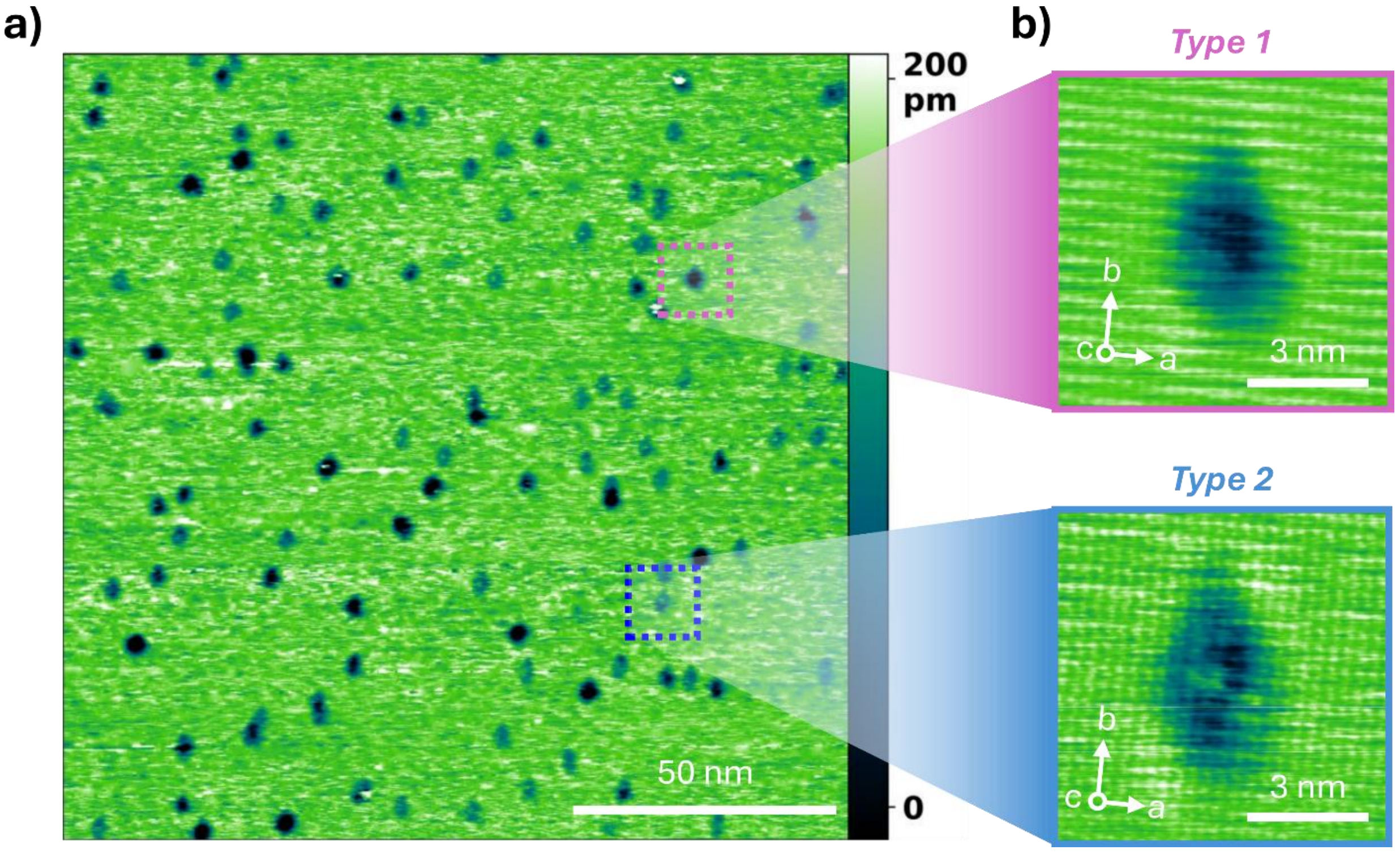

**Figure 2.** a) Typical STM topographic map ($V_B$ = -0.55 V, $I_S$ = -50 pA). Two primary defect types are observed, outlined in purple and blue dashed squares respectively. b) Higher magnification STM topographic maps of the two defect types ($V_B$ = -0.55 V, $I_S$ = -50 pA). All measurements were taken at 211K.

Atomically resolved topographic images on these individual defects (Figure 2b) reveal their qualitative differences: the type 1 defect appears primarily dark, indicating a reduction in the density of states between the scanning bias and charge neutrality, which decays spatially over 1.5 ± 0.1 nm and 2.3 ± 0.1 nm along the a and b axes respectively, corresponding to approximately 4.5 lattice constants in both directions. By comparison, while the type 2 defect features a comparable reduction in states and spatial attenuation, it also displays a relatively sharp increase in contrast near its center, serving as a differentiator from the first defect type. Comparing the symmetry of both defects to the underlying crystal structure, we find that they reflect the two-fold rotational symmetry of the CrSBr surface.

In Figure 3, we explore their electronic signatures through bias-dependent STM topographs and corresponding differential conductance (dI/dV) maps for both defect types.

For the type 1 defect, we do not observe any significant change to the spatial extent of the defect in the bias-dependent topography (Fig. 3a, upper panels). However, upon approaching the edge of the valence band, the defect contrast decreases. We attribute this change to increased tunneling to bulk states of CrSBr, reducing the overall signal-to-noise ratio of the defect states that provide the topographic contrast. This agrees with the increase in the density of states observed near -1.15 V in Fig. 1c, which can obscure signatures of defects in the valence band. Similarly, the defect's signature in the dI/dV maps displays a bright feature centered on the defect that fades lightly towards higher energies. We acquired a similar series of topographic and dI/dV maps for the type 2 defect, presented in Figure 3b. The progressive wash-out of the defect in both sets of maps is observed at comparable energies, however we notice a few key differences compared to type 1. Focusing on the topography panels, we observe that the dim center of the defect becomes more pronounced and splits into two separate points starting around -600 meV.

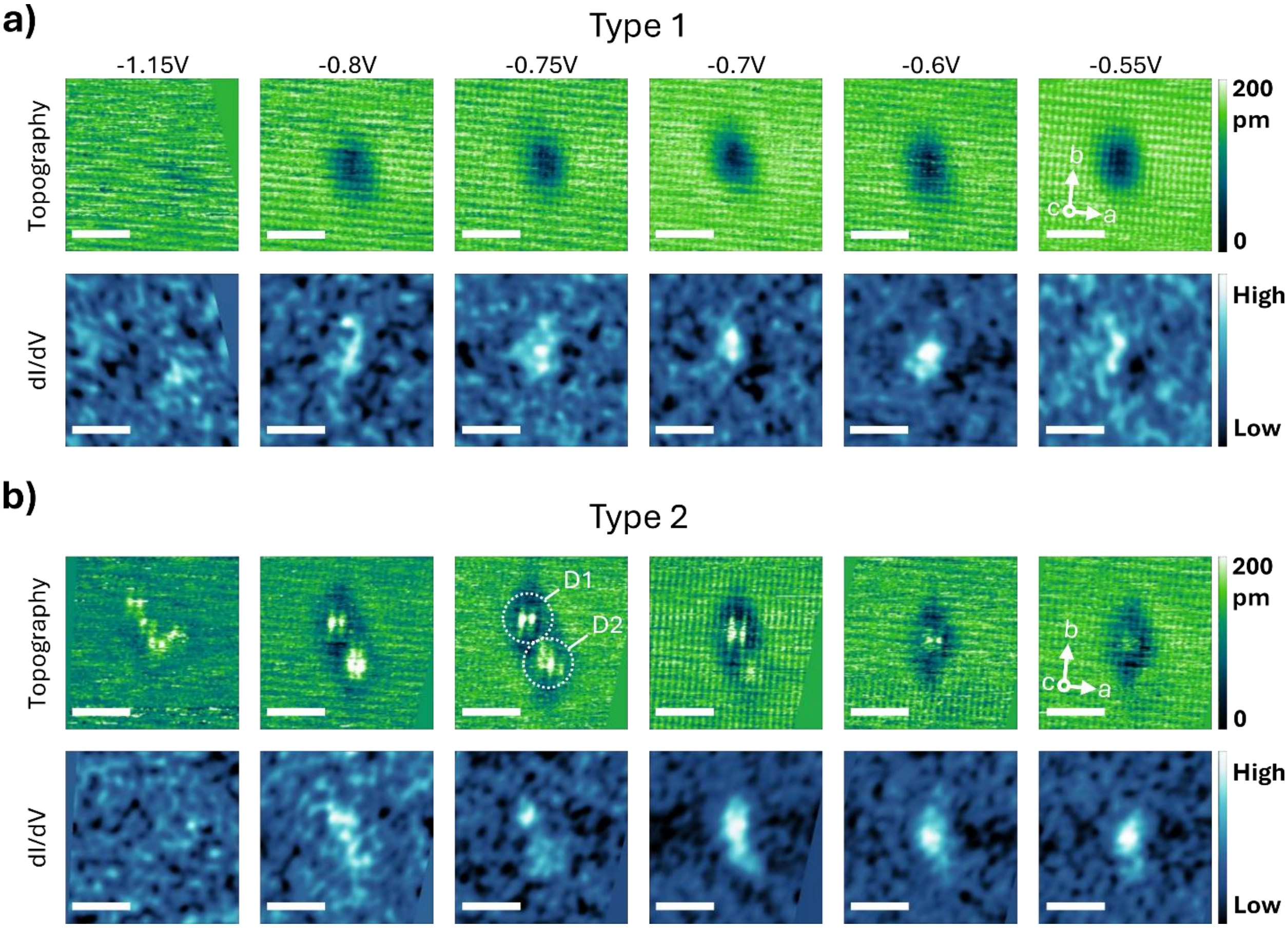

**Figure 3.** Series of STM topographs and corresponding dI/dV maps acquired from -0.55V to -1.15V, as indicated, showing the evolution of the type 1 (a)), and type 2 (b)) defects near the valence band edge. (IS = -50 pA). All scale bars are 3nm. All measurements were taken at 211K.

Moving further towards the valence band edge, we observe that these two central points become elongated along the highly dispersive axis (b-axis) of CrSBr and feature sharper extremities, most evident near -700 meV. In addition, at this energy, we observe that the bright features oriented along the b-axis, corresponding to the defect, appear closer to each other as compared to the atomic lattice, and the column of atoms in between them fades.

Near -750 meV, the elongated bright feature becomes point-like, with faint tails along the b-axis. At this same energy, we start to observe the appearance of an additional defect (D2) on the lower right side of the first (D1) which becomes more apparent at -800 meV. This new defect appears similar to D1, but the two bright points appear closer together, with a stronger contrast. When decreasing the bias voltage further (-1150 meV), we start to lose resolution of the defect. Simultaneously, we observe the formation of additional pairs of bright points on the upper right and upper left of D2, suggesting the formation of additional defects in close proximity.[24,30,31] The formation dynamics of the new defects continues as we ramp the bias voltage back down to -550 meV, as shown in Supplemental Figure S1.

In the differential conductance maps (Figure 3b, bottom panels), the type 2 defect also exhibits a localized electronic state centered on the defect, which follows the symmetry of the lattice and gradually fades at higher energies. We note that the newly formed defects likewise give rise to bright features in the dI/dV maps.

To elucidate the nature of these defect types, we performed *ab initio* DFT+U calculations to obtain band structures and simulated STM maps for a variety of possible defects in CrSBr, which can be found in Supplemental Figure S2. Comparisons between STM measurements and simulations allow us to rule out several candidate defects.

Among the possibilities explored, we found one particular defect in good qualitative agreement with the experimentally observed characteristics of the type 2 defect, as shown in Figure 4.

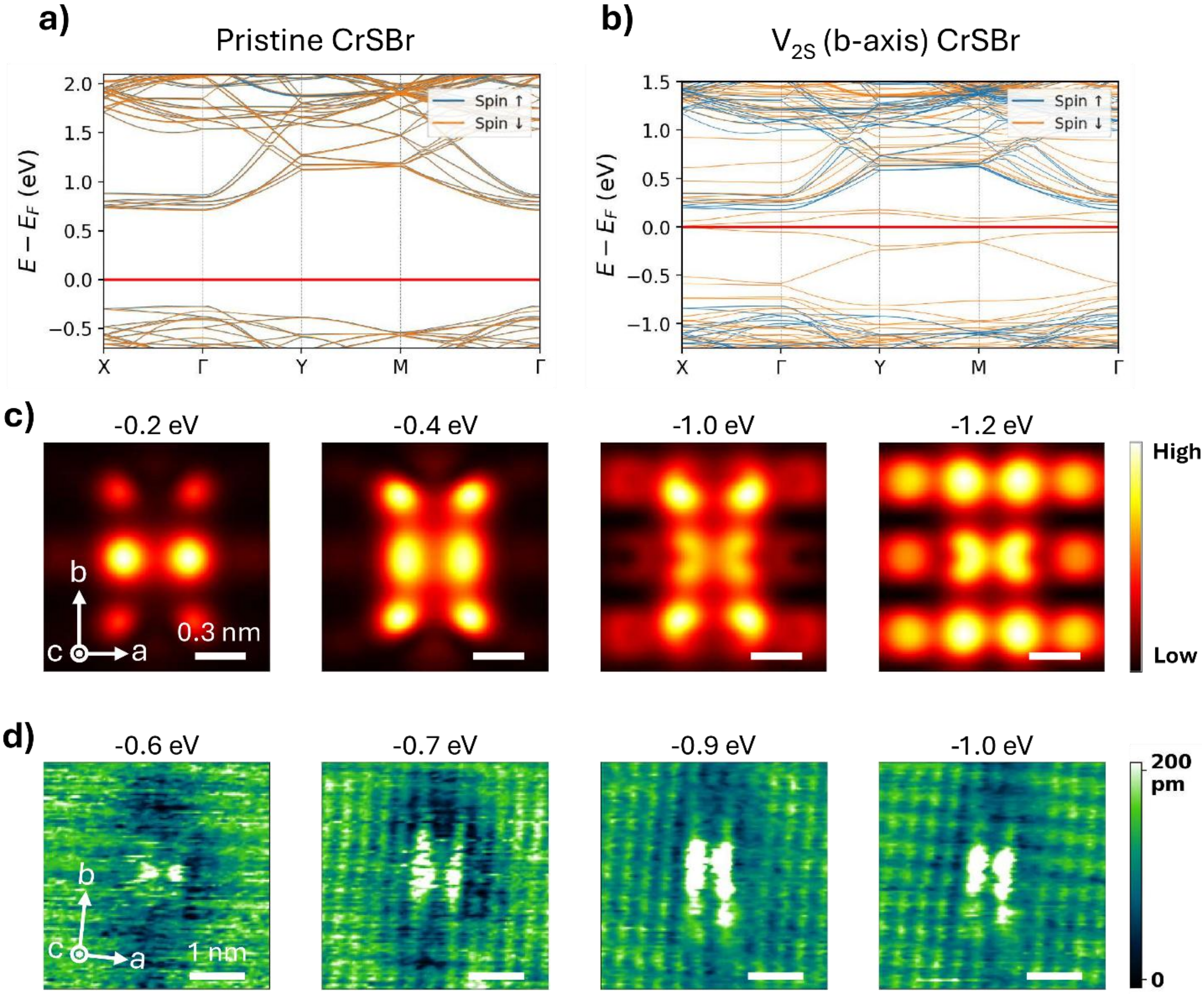

**Figure 4.** a) DFT + U calculated band structure of pristine bilayer CrSBr. b) Calculated band structure of bilayer CrSBr including a double sulfur vacancy aligned along the b-axis. For panels (a) and (b), blue and orange bands are spin up and down respectively and the Fermi energy is marked by a red line. c) Simulated STM maps at the indicated energies centered on a $V_{2S}$ (b-axis) defect in bilayer CrSBr, where the defect is located in the top-most layer. d) STM topographic images centered on the type 2 defect, aligned to highlight shared features with panel (c).

In Figure 4a we show the band structure of the pristine 4x3 bilayer CrSBr supercell, for reference. We see the expected quasi-one-dimensional electronic behavior in the lowest conduction bands,[9,10] with the dispersion along the Γ-X (a-axis in real space) direction being relatively flat, and along the Γ-Y (b-axis in real space) direction being highly dispersive. Inspired by the results of Tulchinsky et al.,[25] which show that defect concentrations in CrSBr can be strongly reduced by increasing the abundance of sulfur during the synthesis process, we focus on sulfur-defect complexes. In Figure 4b we present the calculated band structure for a bilayer CrSBr supercell with a double Sulfur vacancy ($V_{2S}$), aligned along the b-axis in the surface layer. This calculation reveals three features which differentiate the system from its pristine counterpart. Firstly, deep into the conduction and valence bands, most states are no longer spin-degenerate. Secondly, the defect features spin-polarized in-gap bands with large effective mass along the Γ-X direction and lighter in the Γ-Y direction following the one-dimensional behavior of the pristine material. Lastly, the DFT Fermi level lies near the conduction band edge. This suggests n-doping behavior resulting from the defects and is in line with previous reporting.[32] While this might be read as a discrepancy with the experimentally observed Fermi level, (Fig. 1c, bias voltage = 0.0V) near the center of the CrSBr band gap, we note that the Fermi level obtained in such calculations depends decisively on the size of the supercell. Here a 4x3x2 supercell is used containing one $V_{2S}$ defect, corresponding to an effective defect concentration of $5*10^{13}$ $cm^{-2}$, approximately two orders of magnitude above what is observed experimentally. This elevated defect concentration also leads to an exaggerated dispersion of the in-gap defect states. We also note that our calculations are limited by the fact that quantitative predictions of bandwidths, -energies, and -gaps in DFT(+U) are known to be (partially poor) approximations for weakly screened semiconductors.

For qualitative comparison to experiment, we simulated STM images for the $V_{2S}$ defect as described in the Supplemental Information. These results are presented in Figure 4c at the indicated energies.

Starting within the valence band at -1.2 eV, the calculations display bright points that coincide with the positions of the surface-most Bromine atoms of CrSBr with a small increase in intensity when approaching the defect, shown as the central two columns. The two central bright points of this defect appear attracted towards each other (along the a-axis). At higher energy (-1.0 eV), the contrast on the defect increases, the top and bottom most bright points appear lightly rounded, and the two central intensities retain the attraction-like feature. Within the gap (-0.4 eV), the resulting map has increased contrast, where the upper and lower corners of the bright defect appear sharper, with pronounced central points. Finally, near the Fermi level (-0.2 eV), the feature appears primarily as two bright points, and all former features are minimally present.

Comparing this series of calculated STM maps with our experimental data, we observe good qualitative agreement with the STM topographic images taken on the type 2 defect, as detailed in the corresponding maps of Figure 4d. At higher energies, both maps feature the same two central bright points aligned along the a-axis of CrSBr. Moving towards the valence band, these bright features are elongated along the b-axis, becoming broader and appearing closer to one another.

The qualitative resemblance between our STM measurements and simulation suggests that $V_{2S}$ is a likely candidate impurity complex responsible for the observed electronic characteristics of the type 2 defect. We note that the STM/STS measurements do not resolve sufficient internal structure for the type 1 defect to enable its conclusive identification by comparison with the simulated defect structures, revealing only a local reduction of the density of states.

In conclusion, our STM investigation of CrSBr revealed two dominant defect species, featuring distinct spatial and electronic signatures near the valence band edge. These defects preserve the anisotropic symmetry of the CrSBr lattice and exhibit changes to the local density of states within the CrSBr band gap. Bias dependent measurements further demonstrate that these defects possess strong energy-dependent structure. The type 2 defect exhibits atomic-scale structure that enables direct comparison with DFT+U simulations, identifying it as consistent with a b-axis-aligned double sulfur vacancy. This defect complex imprints an interesting electron impurity structure into the band gap of the CrSBr host, specifically in the form of energetically close-lying occupied and unoccupied states of the same spin species. Consequently, this $V_{2S}$ could enable single-photon emission at low energies or long wavelengths that is tunable by a magnetic field.

These findings establish a direct comparison between defect configurations and electronic signatures in CrSBr, providing an important step towards understanding how defects influence the electronic properties of this layered semiconductor. Given the strong coupling between the electronic structure and magnetism in CrSBr, identifying and controlling such defects are a critical step towards defect-engineering strategies for tuning excitonic responses, magnetic phases, and low-dimensional spintronic functionalities in this emerging class of quantum materials.

AUTHOR INFORMATION

**Corresponding Author**

*aluica2@uic.edu

**Author Contributions**

A.L.-M. conceived and supervised the project. A.S. synthesized the CrSBr crystals under the supervision of Z.S. J.B., N.G., and K.J. prepared the STM sample. J.B. and N.G. performed the STM/STS measurements. N.G. provided technical support for the STM experiment. J.B. analyzed the STM/STS data and wrote the initial draft of the manuscript, with input from all authors. J.A. performed the theoretical modeling under the supervision of M.R. All authors contributed to the interpretation of the results and reviewed and approved the final version of the manuscript.

## METHODS

### Experimental details

STM measurements were performed using a Scienta Omicron Inc. Variable Temperature AFM/STM with a SPECS Group Nanonis STM Controller. All measurements were taken at 211K in UHV using an electrochemically etched Tungsten tip. Bulk CrSBr was mounted onto a steel sample plate using conductive vacuum-safe epoxy and cleaved in-vacuum to obtain a pristine CrSBr surface.

dI/dV maps were obtained simultaneously with topographic images using a lock-in amplifier and a bias modulation of 10mV. All STM images shown in this work were corrected for drift by tracking the movement of features between successive scans and applying the corresponding affine transformation. The resulting images were validated by confirming the expected periodicity and symmetry of the CrSBr atomic lattice.

The CrSBr band gap was obtained by fitting linear curves to the conduction and valence band onsets of 8000 spectra. The point where these curves intersect the minimum of the dI/dV were marked as the band edge.

### Theoretical details

DFT+U calculations were performed using the Perdew-Burke-Ernzerhof (PBE) GGA exchange functional[33] with van der Waals corrections according to DFT-3 method of Grimme within a basis of projector augmented waves.[34,35] The local coulomb interaction in DFT+U was included following ref[36] as implemented in Vienna *ab initio* simulation package (VASP) version 6.4.2,[37,38] with parameters U=2.5 eV and J = 0.4 eV. The charge density was determined self-consistently on k-grid of 3x3x1, with a plane wave cut-off energy of 500 eV and Gaussian smearing of 0.05 eV. The electronic self-consistency condition was set to $10^{-7}$eV.

The structures used were 4x3x2 supercells of the primitive bulk CrSBr unit cell. A vacuum of about 16 Å was added to simulate STM at the surface. Desired impurity complexes were constructed by removing atoms. Then the structures were relaxed until forces on the ions were smaller than $5 * 10^{-5}$eV/ Å. Relaxations were done at with one k-point at Gamma.

STM simulations were performed in the Tersoff-Haman approximation[39] for a point-like tip at a constant height of 3 Å above the outermost Br that allows tunneling to all electronic states within an energy window from the Fermi-level down to the applied bias voltage with equal probability.

ACKNOWLEDGMENTS

Work performed at the Center for Nanoscale Materials, a U.S. Department of Energy Office of Science User Facility, was supported by the U.S. DOE, Office of Basic Energy Sciences, under Contract No. DE-AC02-06CH11357. J. A., and M. R. acknowledge support from the research program "Materials for the Quantum Age" (QuMat). This program (registration No. 024.005.006) is part of the Gravitation program financed by the Dutch Ministry of Education, Culture and Science (OCW). M.R. acknowledges support from the Vidi ENW research programme of the Dutch Research Council (NWO) under Grant DOI: 10.61686/YDRHT18202 (file number VI.Vidi.233.077). We thank Shen van Hassel and Sergii Grytsiuk for supporting the structural relaxation calculations. Z.S. was supported by ERC-CZ program (project LL2101) from Ministry of Education Youth and Sports (MEYS) and by project LUAUS25268 from Ministry of Education

Youth and Sports (MEYS). J.B. acknowledges support from the Natural Sciences and Engineering Research Council of Canada (NSERC) ALLRP/578466-2022. A.L.-M. was supported in part from CIFAR and the Gordon and Betty Moore Foundation Experimental Physics Investigators program and the US Department of Energy, Office of Science, Basic Energy Sciences, Materials Sciences and Engineering Division under contract DE-AC02-06CH11357.

SUPPORTING INFORMATION

**Supporting Information Available:** Supplemental Figures 1 and 2 (PDF)

# Visualization of Defect Electronic States in Layered Semiconductor CrSBr (Supplemental Information)

**Supplemental Figures:**

Figure S1 shows a series of STM topographic images of the type 2 defect, labeled in the order of acquisition. As the sample bias is decreased, the secondary defect D2 begins to emerge, as discussed in the main text. However, upon increasing the bias back toward the Fermi level, the defect undergoes a permanent change in appearance. This change is most clearly seen by comparing panels 7 and 8, or panels 1 and 11, where the dark depression initially surrounding D1 has migrated to D2. This behavior indicates that the electronic configuration of the defect has been modified during the measurement.

Figure S2 presents the calculated band structures, simulated STM images, and relaxed crystal structures for the defect configurations considered in this work. These include double sulfur vacancies aligned along the b axis and a axis, single sulfur vacancies, and sulfur–bromine vacancies located in both the upper and lower chalcogen–halogen layers of CrSBr. Among the defect configurations investigated, only the b-axis-aligned double sulfur vacancy ($V_{2S}$) exhibits qualitative agreement with the experimental STM observations. Although the spatial signatures differ substantially between the simulated defects, all configurations similarly exhibit a tendency toward *n*-type doping.

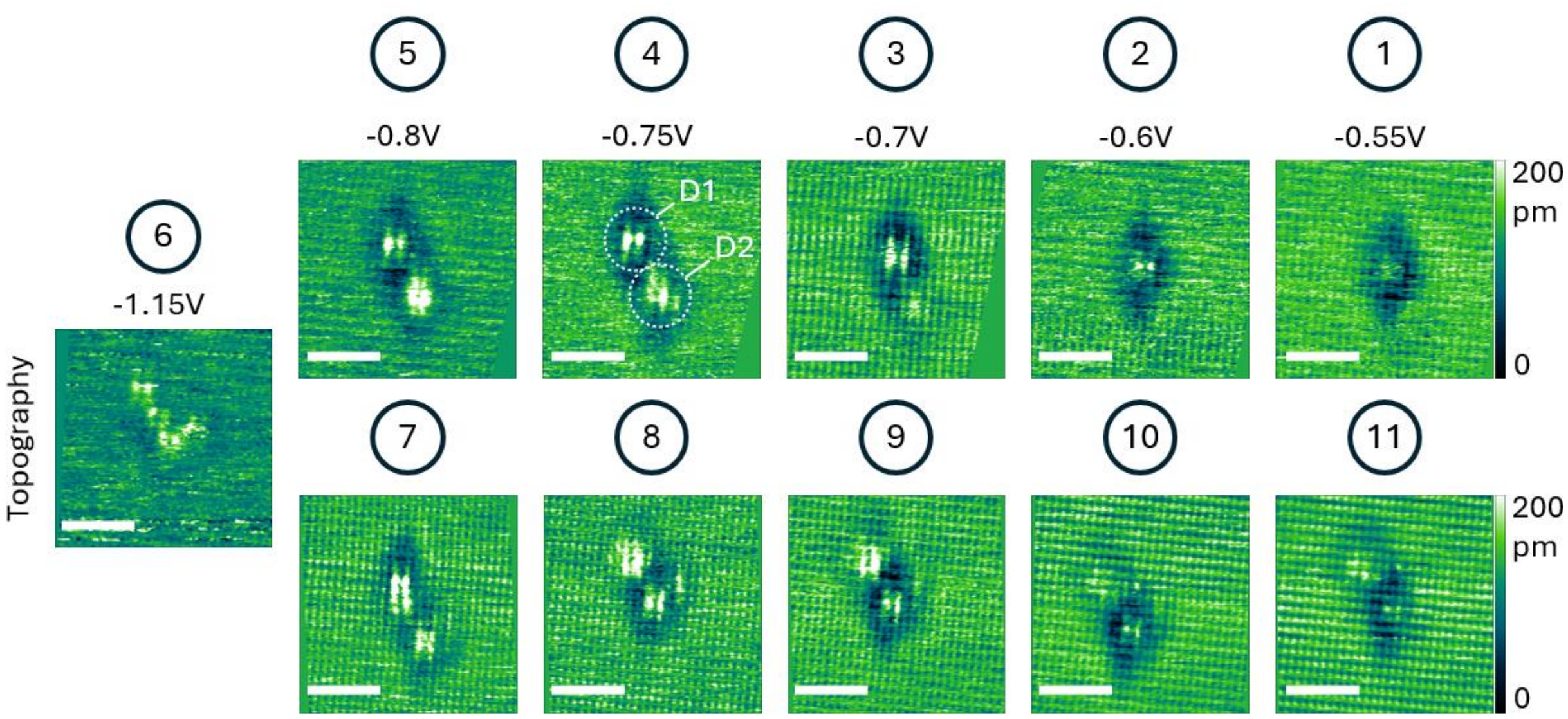

**Figure S1.** Series of STM topographic images taken from -0.55V to -1.15V to -0.55V illustrating the evolution in the type 2 defect states. The number above individual images indicated the order in which the scans were collected. The dark cloud around D1 appears to migrate to D2 when transitioning between scan 7 to scan 8. All scale bars are 3nm.

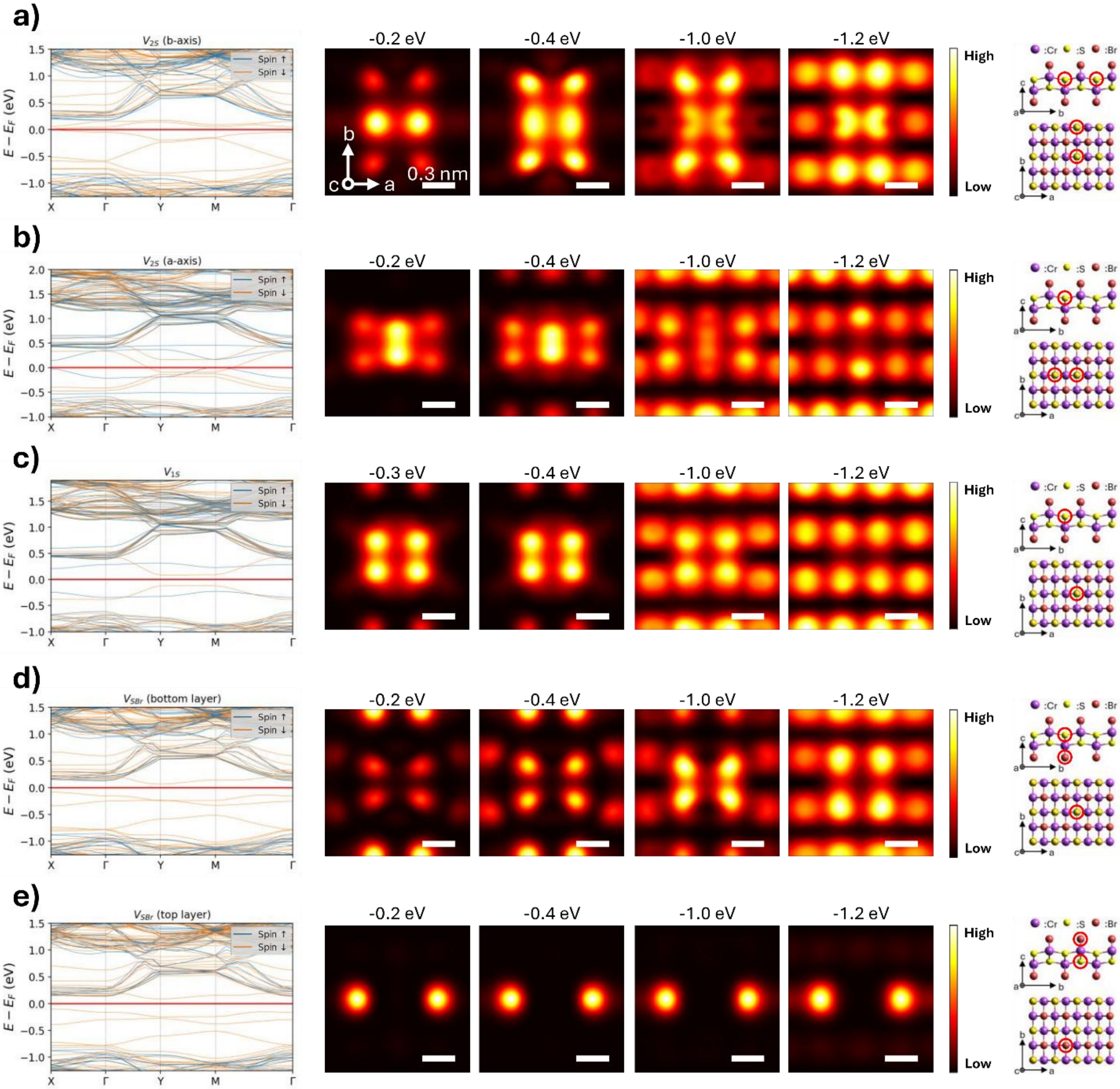

**Figure S2.** Calculated band structures, simulated STM maps and crystal structures for defects considered in this work. a) Band structure and simulated STM maps for a ( (a) double sulfur vacancy aligned along the b-axis, (b) double sulfur vacancy aligned along the a-axis, (c) single sulfur vacancy, (d) sulfur and bromine vacancy in the bottom row, and (e) sulfur and bromine vacancy in the top row.